# Universal Routing of Light via Optical Thermodynamics

Hediyeh M. Dinani[1,*], Georgios G. Pyrialakos[1,*], Abraham M. Berman Bradley[1], Monika Monika[2], Huizhong Ren[1], Mahmoud A. Selim[1], Ulf Peschel[2], Demetrios N. Christodoulides[1,3], and Mercedeh Khajavikhan[1,3]

[1] Ming Hsieh Department of Electrical and Computer Engineering, University of Southern California, Los Angeles, CA 90089, USA.
[2] Abbe Center of Photonics, Friedrich Schiller University Jena, 07743 Jena, Germany.
[3] Department of Physics and Astronomy, University of Southern California, CA 90089, USA
[*] These authors contributed equally

Corresponding authors: khajavik@usc.edu (M.K.) and demetri@usc.edu (D.N.C.)

Understanding and exploiting the dynamics of complex nonlinear systems is nowadays at the core of a broad range of scientific and technological endeavors. Within the optical domain, light evolution in a nonlinear multimode environment presents a formidable problem, as its chaotic evolution often hinders predictive insights. Recently, an optical thermodynamic framework has been put forward that, in a systematic manner, can not only predict but also harness the intricate behavior of these systems. In this work, by deploying entropic principles, we demonstrate a counterintuitive optical process in which light, launched into any input port of a judiciously designed nonlinear array, universally channels into a tightly localized ground state – a response that is completely unattainable in linear conservative arrangements. This phenomenon arises from the interplay between lattice structure and the way the kinetic and nonlinear Hamiltonian components unfold, leading to two optical thermal processes—a Joule-Thomson-like expansion followed by mode thermalization. Experimentally, this effect is demonstrated in properly configured nonlinear time-synthetic mesh lattices, where the optical temperature approaches near zero, causing light to condense at a single spot, regardless of the initial excitation position. The effect demonstrated here opens new avenues for applying the principles of optical thermodynamics in realizing novel optical functionalities, such as all-optical beam steering, multiplexing, and nonlinear beam shaping in high-power regimes, while also offering a greater understanding of the remarkable physics of light-matter interactions in multimode nonlinear systems.

Chaotic dynamics are a defining feature of complex systems across various fields, ranging from turbulent particle interactions in gases [1, 2], to phase transitions in magnetic materials [3–5], and protein folding in biological systems [6, 7]. Multimode optical arrangements present a similarly intricate landscape, where nonlinearities intertwine multiple degrees of freedom, leading to complex and seemingly unpredictable wave dynamics [8–12]. Despite their perplexing nature, these photonic systems hold promise for revealing new, uncharted behaviors, suggesting realms previously considered beyond reach. Lately, a self-consistent optical thermodynamic framework grounded in entropic principles [13–15] has been developed to address these complexities. Just as classical thermodynamics [16] has driven transformative advancements—enabling the design of heat engines, the prediction of phase transitions [17], and the understanding of energy flow in biological and cosmological systems [18]—this methodology provides similar predictive capabilities for complex light dynamics through universal laws and macroscopic parameters such as optical temperature and entropy [19–31].

In this work, we leverage tenets from statistical physics to demonstrate, for the first time, efficient and universal light routing in conservative nonlinear waveguide arrays. This process allows light to reliably reach a designated output channel, regardless of its entry port to the array (Figs. 1a, b). We note that under linear conditions, this functionality is unattainable without employing significant levels of gain, especially as the number of input ports increases [32, 33]. In other words, it is impossible to conceive of a unitary transformation that enables efficient light transport from a multitude of input sites (when excited one at a time) to a preassigned exit port, as this would violate reciprocity [33]. Nonlinearity could perhaps overcome this fundamental hurdle; however, the appropriate strategy to achieve this goal remains unclear. As we will see, this elusive funneling capability can be achieved through a Joule-Thomson exchange [29] between the photonic kinetic and nonlinear Hamiltonian components of the system, resulting in rapid cooling of light to a near-zero optical temperature, followed by mode thermalization. This effect emerges in conservative potential landscapes that favor localized lower-order states on one side of the lattice while supporting a set of extended modes in the bulk. Clearly, the realization of universal funneling methodologies could enhance the arsenal of tools available in photonics, which are aimed at controlling and manipulating the flow of light [34–42]. In what follows, we will experimentally demonstrate this behavior using a nonlinear photonic time-synthetic optical mesh lattice platform [19, 32] (Figs. 1c, d), which will be discussed in greater detail subsequently.

To exemplify this intriguing prospect, we perform simulations in a prototypical one-dimensional discrete Kerr nonlinear array [43], with triangular on-site energies. This lattice closely mirrors the mesh-lattice environment used in our experiments (Fig. 2a). In this arrangement, the complex optical field state vector $|\Psi\rangle$ evolves according to $i\frac{d|\Psi\rangle}{dz} + (\widehat{H}_L + \widehat{H}_{NL})|\Psi\rangle = 0$, where $\widehat{H}_L$ is a normalized Hermitian matrix comprising off-diagonal nearest-neighbor coupling elements as well as local (diagonal) detunings or energies $\Delta_n$, while the diagonal operator $\widehat{H}_{NL}$ accounts for the Kerr nonlinearity. In general, the state vector of $|\Psi\rangle$ can be projected on the linear supermodes $|u_i\rangle$ of the system ($\widehat{H}_L |u_i\rangle = \varepsilon_i |u_i\rangle$) having eigenvalues $\varepsilon_i$, in which case $|\Psi\rangle = \sum_i c_i |u_i\rangle$ where $c_i$ represents the respective complex mode coefficient. In conservative optical multimode settings, nonlinearity induces a chaotic and, thus, ergodic energy exchange among the modal occupancies $|c_i(z)|^2$—an effect that underlies their thermodynamic response. These statistical processes are governed by two constants of motion: the total Hamiltonian energy $H_{tot} = H_L + H_{NL}$, which involves both a linear and a nonlinear component, and the total optical power $P = \sum_i |c_i(z)|^2$. The linear part of the Hamiltonian $H_L = -U = \sum_i \varepsilon_i |c_i(z)|^2$ is associated with the "kinetic" energy of the system, whereas the nonlinear contribution, $H_{NL} = 1/2 \sum_n |a_n(z)|^4$ [43], (expressed in the local basis $|\Psi\rangle = (a_1, \ldots, a_n)$), signifies a nonlinear interaction energy. To facilitate a funneling path through thermodynamics, the array is designed with highly localized lowest-order modes confined to one side, while higher-order (bulk) modes remain extended throughout the lattice. This is achieved by progressively adjusting the detuning between individual sites at a constant rate $\delta\Delta$ ($\delta\Delta = \Delta_n - \Delta_{n-1}$), thereby forming a truncated triangular discrete potential in this 1D arrangement (Fig. 2a). The same methodology applies to two-dimensional lattices, where the local energies assume a conical-like profile (Fig. 1b).

The thermal aspects of this funneling process are elucidated in Fig. 2b, where a high-intensity signal or beam is injected into this triangular potential lattice. From an energetic perspective, initially, the system is dominated by a nonlinear Hamiltonian component $H_{NL}$, that happens to be in a superposition of many higher-order modes. As the light packet evolves in the lattice, it gradually sheds some energy because of Peierls-Nabarro effects [44], thus forming a moderately confined, discrete soliton-like entity [43, 45] that, like a particle, continuously accelerates (Fig. 2b) within the linear (biased) potential lattice [49]. As a result, the modal content of the soliton beam progressively changes (Fig. 2c) while its kinetic energy $|U|$ steadily increases at the expense of the nonlinear component (Fig. 2d). We note that our lattice is deliberately designed to display a small step $\delta\Delta$ to suppress Bloch oscillations [46, 47] through the action of nonlinearity. Eventually, this soliton state reaches the end of the triangular array where the lowest-order mode resides. Here, the abrupt collision leads to a series of reflections, causing the beam to completely disintegrate into a low-temperature 'gas' state, where the Hamiltonian energy is almost entirely converted into kinetic $U$, which remains quasi-invariant thereafter. Beyond this point, the photon gas enters a second weakly nonlinear phase where it attains thermal equilibrium, as revealed by the emergence of a Rayleigh-Jeans (RJ) distribution (Fig. 2e). At this stage, the expectation value of the modal content is given by $\langle |c_i|^2 \rangle = -\frac{T}{\varepsilon_i + \mu}$, where $T$ is the optical temperature, and $\mu$ the chemical potential [13, 20–22, 28]. This all-optical energy transformation process, akin to Joule-Thomson (JT) expansion [29] encountered in standard statistical mechanics [16], is the driving mechanism for the light funneling discussed in this work. For the example provided here, the final temperature is exceedingly low ($T = 0.012$), suggesting indeed that the optical energy predominantly occupies the ground state (Fig. 2e). Importantly, this effect is universal, as the optical energy is faithfully funneled into the system's highly localized fundamental mode regardless of the input excitation site.

To experimentally demonstrate universal light funneling in a nonlinear lattice environment, we designed a nonlinear fiber loop setup (Fig. 1c) that allows one to observe the evolution of light packets in discrete time steps [19] (Suppl. I). Our setup consists of two ~3 km long dispersion-compensating fiber loops exhibiting an appreciable Kerr nonlinearity, that are slightly unequal in length (~ 20 m). These loops are coupled together by a variable coupler (VC), effectively establishing a 1D lattice with the same nearest neighbor coupling (Suppl. II). A 20 ns pulse from a highly coherent DFB laser is subsequently injected into one of the loops through an optical switch (SW) (Fig. 1c). Using a time-multiplexing scheme, pulse sequences traveling in the short ($u$) and long ($v$) loops are temporally advanced and delayed, respectively, resulting in discrete time slots (synthetic positions $n$) within each round trip (time step $m$), as shown in Fig. 1d. During this process, the pulses accumulate a nonlinear phase [48] while the use of two phase modulators (PMs), one in each loop, allows the realization of a triangular potential at each site $n$ that reaches a height of $0.324\pi$. Given that the lattice involves 52 sites (or supermodes), the site-to-site phase difference is $\delta\phi = \phi(n, m) - \phi(n - 1, m) = 0.0062\pi$.

In our experiments, funneling was observed when the injected pulse peak power was around ~160 mW and after approximately $m = 100$ time steps. Figure 3a illustrates the experimentally observed light evolution in a triangular potential lattice when a signal is injected at the $n = -10$ site of the short loop ($u$). In line with the preceding theoretical discussion, as the pulse traverses the array, it initially forms a soliton-like state after some readjustment. This self-confined state then accelerates, as indicated by its parabolic trajectory (Fig. 3a), within the time interval $0 \leq m \lesssim 80$, eventually striking the triangular barrier and undergoing a series of self-bouncing reflections lasting up to $m \sim 125$. This initial cooling phase is characterized by two crucial aspects: (i) a continuous shift in the eigenfunctions constituting the beam, ultimately engaging primarily the fundamental mode located at the peak of the triangular potential, and (ii) an increase in the kinetic energy component $U$ (Fig. 3c), which progressively approaches the value $U \sim -\varepsilon_1 P_1$, where $\varepsilon_1, P_1$ represent the eigenvalue and power associated with the ground state, respectively [29]. The behavior displayed in Fig. 3a is in excellent agreement with numerical simulations (Fig. 3b). Beyond $m \sim 100$, the soliton fragments undergo JT thermalization, resulting into a cooled

Rayleigh-Jeans distribution (Fig. 3d) having a normalized optical temperature of $T = 0.004$ and a chemical potential $\mu = -2.38$, indicating that most of the power occupies the fundamental mode. To demonstrate the universality of this process, we further examine the system's behavior under excitation from different sites. The experimental results for input locations $n = -20$ and $n = -30$ in the short ($u$) loop, along with the corresponding light evolution simulations, are presented in Fig. 3e and Fig. 3f, respectively. When the injection site is further from the center, higher-order modes, which tend to spread toward the edges of the lattice, are preferentially excited. This results in a slightly slower progression of the optical cooling process. Here, the transition from JT expansion to optical thermalization occurs at around $m\sim 120$, as the system requires more time to redistribute its energy toward the lower-order modes near the peak of the triangular potential. Nonetheless, in both cases, the system guides light toward the fundamental mode as expected from a universal router. Additional results, along with findings that further confirm the universality of this process, are provided in the Supplementary Information Note VI and IX.

Light funneling, as described above, arises from the intricate dynamics of moderately nonlinear multimode systems, whose behavior markedly differs from both linear arrays and highly nonlinear scenarios. As a result, for a given potential landscape, universal routing occurs within a certain range of optical powers. At low power levels, the system operates in a quasi-linear regime that in principle, in large arrays, can display weakly interacting Bloch oscillations [46, 47]. These oscillations tend to disperse light across multiple sites, thus preventing the emergence of power-siphoning behavior (Fig. 4a, first panel). As the injected power increases, nonlinear phase accumulation takes over, gradually suppressing these oscillations and establishing a moderately confined accelerating soliton state. This, in turn, enables a funneling regime where light is efficiently directed toward the designated output site via JT cooling and mode equilibration (Fig. 4a, middle panel). Across this power range, universal routing is consistently observed in all trials, demonstrating the robustness of this mechanism. At higher power levels, the system enters a strongly nonlinear regime where strong soliton self-focusing overrides the potential's influence because of pronounced Peierls-Nabarro effects [44]. In this state, a highly localized discrete soliton [43] forms that chart independent trajectories around the excited site, once again preventing the funneling behavior from occurring (Fig. 4a, right panel). Figure 4b further illustrates the efficiency of universal routing as a function of the array's potential slope and input power, revealing a transition between the solitonic and funneling regimes. It is important to note that the nonlinear dynamics considered here are inherently Liouvillian and subject to time-reversal symmetry. However, any deviation from a perfectly phase-conjugated output is expected to result in significantly different initial conditions at the input, due to the chaotic dynamics unfolding in this complex arrangement. Consequently, while the system is, in principle, time-reversible, the universal routing behavior is expected to be practically irreversible, much like in an actual thermodynamic environment.

In conclusion, we have demonstrated a novel nonlinear light-routing mechanism, where optical signals, regardless of their entry point, are directed to converge at a designated output port. This intriguing behavior arises from the interplay between the array's modal structure and two distinct optical thermodynamic processes enabled by the multimode nonlinear environment: a Joule-Thomson-like expansion, facilitating optical cooling, followed by thermalization, which equilibrates the light into an RJ distribution, ultimately concentrating it into the ground state of the lattice. The principles unveiled here can be extended to other photonic platforms, such as integrated nonlinear waveguide arrays and photonic crystal multicore fibers, where similar funneling schemes could have transformative applications in optical spatial multiplexing systems, beam shaping, and power scaling. Furthermore, our findings highlight the value of optical thermodynamics in unveiling new regimes of light-matter interactions in nonlinear multimode systems, offering fresh insights into the dynamics of complex optical arrangements and advancing their potential for high-power applications.

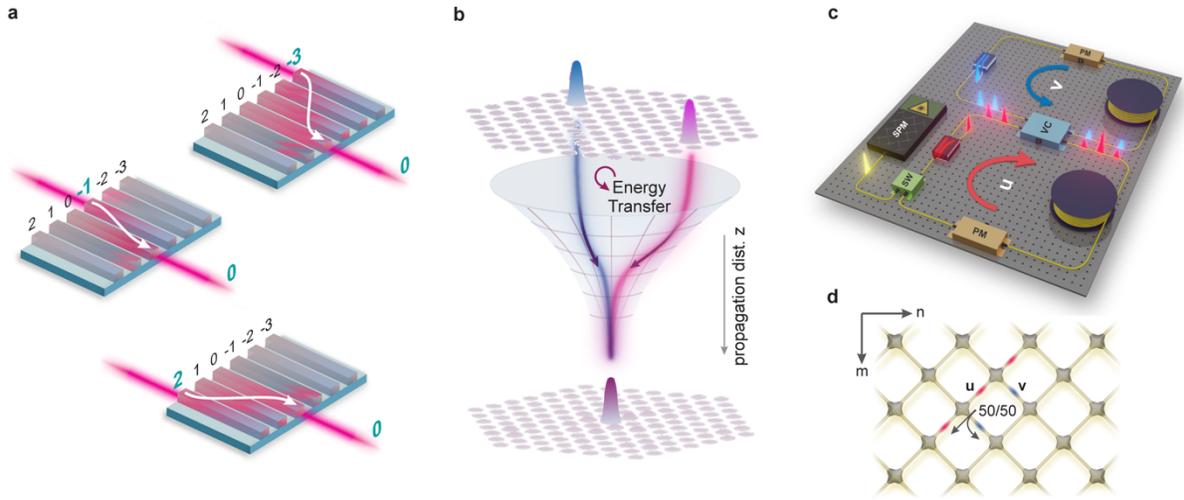

**Figure 1: Nonlinear funneling of light. a**. Conceptual illustration of a photonic integrated array, engineered to universally route light into its central port. This dynamic response is prohibited in linear, conservative systems, even in the presence of nonreciprocity. **b**. Simulation results of funneling in a nonlinear 37-core coupled array, when excited simultaneously at 2 independent ports, with normalized power $P = 10.5$ per port. The lattice exhibits a conical potential with a detuning difference $\Delta_c - \Delta_o = 6$ between the 5 outermost elements ($\Delta_o$) and the central cite ($\Delta_c$). **c**. The simplified experimental setup consists of two nonlinear optical fiber loops of unequal length, connected by a variable coupler (VC). Each loop contains a phase modulator (PM) to control the real part of the lattice potential. The short loop is connected to a pulse source generator, i.e., seed pulse module (SPM), via an optical switch (SW). During propagation, the pulse intensities are monitored through photodetectors. **d**. The pulse propagation dynamics through the short ($u$; red) and long ($v$; blue) fiber loops can be mapped onto a mesh lattice of beam splitters as a function of position $n$ and round trips $m$ (see the 'Generation of Optical Mesh Lattices' section in the Supplementary Materials for details).

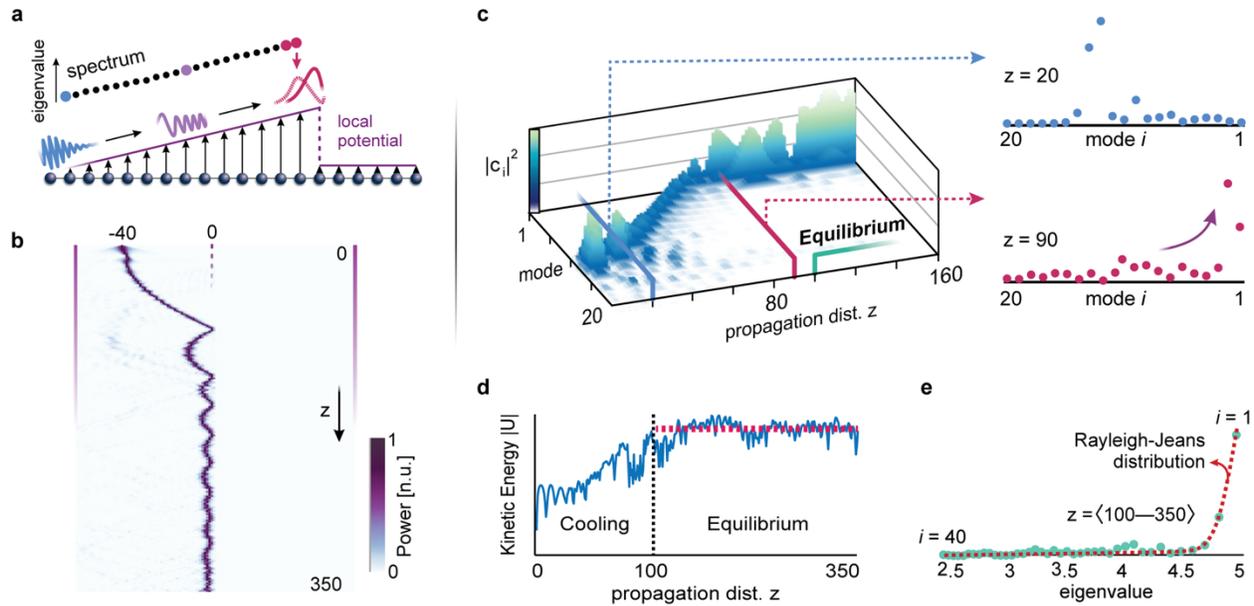

**Figure 2: Thermodynamic principle of light funneling. a**. A coupled array system with its lowest-order modes (red) localized at the peak of its weakly triangular potential. **b**. Within an appropriate range of potential slopes, the funneling regime becomes accessible. Here, a lattice with $\delta\Delta = 0.02$ and peak site at 0 is excited at the port $-40$ with normalized power $P = 4$. **c**. Evolution of modal amplitudes during funneling for the scenario depicted in **b**. The wave packet undergoes optical cooling, progressively carrying power towards the lower-order modes. **d**. During this process, the kinetic energy increases while **e**. a Rayleigh-Jeans distribution manifests at equilibrium (here depicted as a time average of the modal amplitudes for 250 time steps). The theoretical red dashed curve in **e** is calculated using the mean equilibrium value of $U$ (red continuous line in **d**).

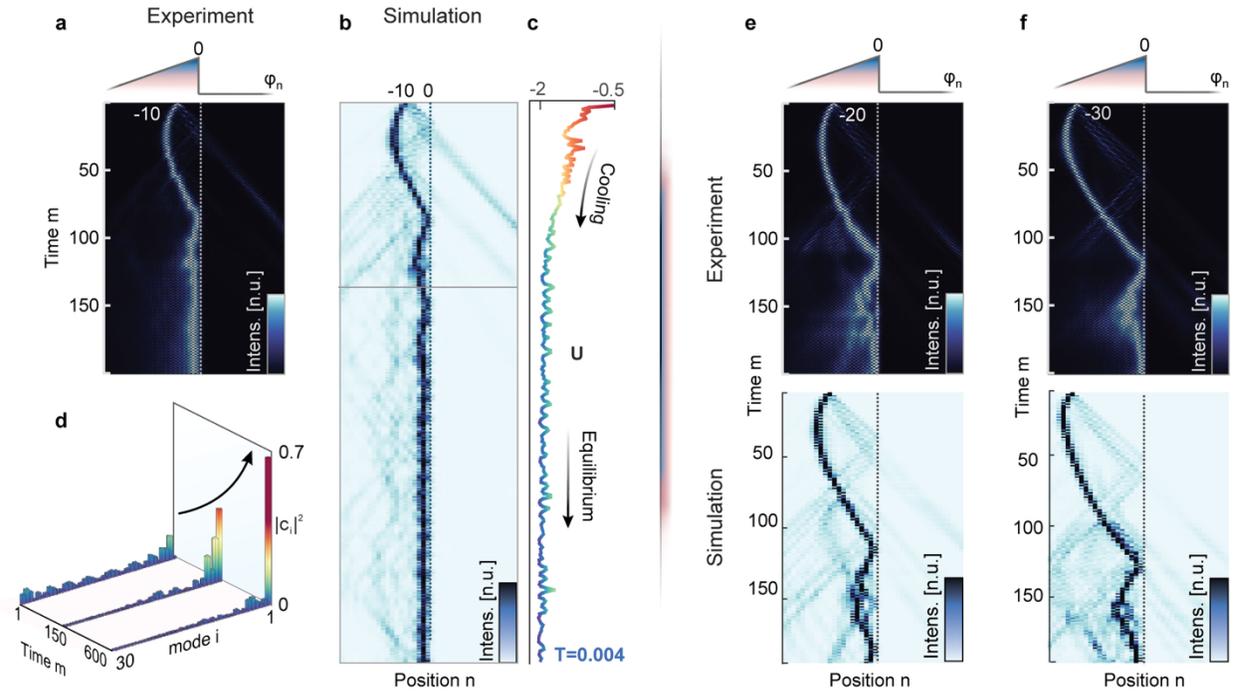

**Figure 3: Experimental observation of universal light funneling**. **a**. Experimental data and **b**. corresponding simulations depict the evolution of a single pulse, when injected at 160 mW at position $n = -10$ in the short loop, within a triangular-shaped potential that abruptly drops at the center of the lattice. **c**. The monotonic increase in the optical kinetic energy $U$ during the propagation indicates that the photon gas has transitioned from "hot" state to a near-zero optical temperature. **d**. The modal occupancy is monitored at three specific time steps ($m = 1$, 150, and 600) by projecting the field onto the lattice supermodes, demonstrating a power transfer toward the fundamental mode. **e, f**. Experimental results concerning the impact the initial injection position has on the funneling process, obtained at $n = -20$ and $n = -30$, respectively. The measured pulse intensities are normalized at each time step.

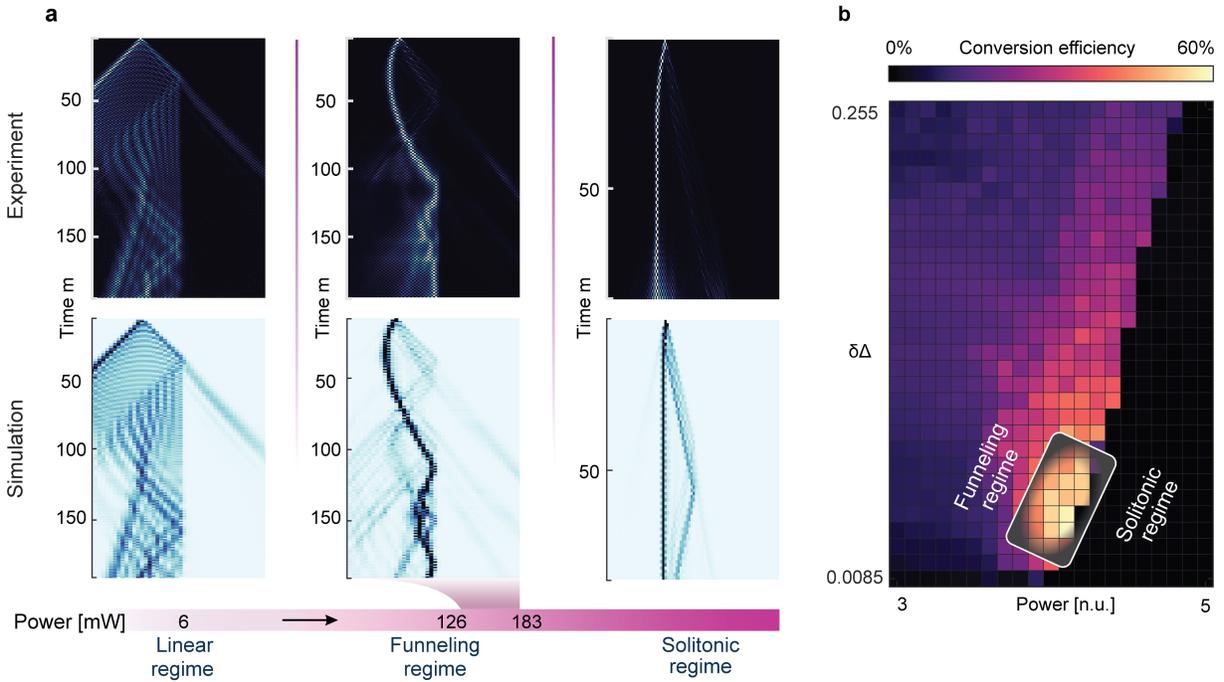

**Figure 4: Regime of optical funneling**. **a**. Under weak nonlinear conditions (left panel), a diffraction pattern is manifested, while at a higher power level, a moderately confined soliton emerges that funnels to a designated output port (middle panel). At higher power levels, a strong soliton is formed that charts its trajectory (right panel). **b**. Funneling efficiency map as a function of normalized power and detuning slope difference δΔ. The region exhibiting efficient funneling is depicted by the inset and borders with the solitonic regime.

# References

[1] Wit, X.M., Fruchart, M., Khain, T., Toschi, F., Vitelli, V.: Pattern formation by turbulent cascades. Nature 627(8004), 515–521 (2024)
[2] Manneville, P.: Instabilities, Chaos and Turbulence vol. 1. World Scientific, (2010)
[3] Pelissetto, A., Vicari, E.: Critical phenomena and renormalization-group theory. Physics Reports 368(6), 549–727 (2002)
[4] Wigen, P.E.: Nonlinear phenomena and chaos in magnetic materials. In: Nonlinear Phenomena and Chaos in Magnetic Materials, pp. 1–12. World Scientific, (1994)
[5] Deng, S., Gomonay, O., Chen, J., Fischer, G., He, L., Wang, C., Huang, Q., Shen, F., Tan, Z., Zhou, R., et al.: Phase transitions associated with magnetic-field induced topological orbital momenta in a non-collinear antiferromagnet. Nature Communications 15(1), 822 (2024)
[6] Uversky, V.N.: Dancing protein clouds: the strange biology and chaotic physics of intrinsically disordered proteins. Journal of Biological Chemistry 291(13), 6681–6688 (2016)
[7] Jumper, J., Evans, R., Pritzel, A., Green, T., Figurnov, M., Ronneberger, O., Tunyasuvunakool, K., Bates, R., Žídek, A., Potapenko, A., et al.: Highly accurate protein structure prediction with AlphaFold. Nature 596(7873), 583–589 (2021)
[8] Boyd, R.W., Gaeta, A.L.: Chaos in nonlinear optics. In: Laser Optics of Condensed Matter: Volume 2 The Physics of Optical Phenomena and Their Use as Probes of Matter, pp. 99–105. Springer, US, Boston, MA, (1991)
[9] Zakharov, V.E., L'vov, V.S., Falkovich, G.: Kolmogorov Spectra of Turbulence I: Wave Turbulence. Springer, (2012)
[10] Mangini, F., Ferraro, M., Tonello, A., Couderc, V., Wabnitz, S.: High-temperature wave thermalization spoils beam self-cleaning in nonlinear multimode grin fibers. Optics Letters 48(18), 4741–4744 (2023)
[11] Longhi, S.: Modulational instability and space time dynamics in nonlinear parabolic-index optical fibers. Optics letters 28(23), 2363–2365 (2003)
[12] Mangini, F., Gervaziev, M., Ferraro, M., Kharenko, D., Zitelli, M., Sun, Y., Couderc, V., Podivilov, E., Babin, S., Wabnitz, S.: Statistical mechanics of beam self-cleaning in grin multimode optical fibers. Optics Express 30(7), 10850–10865 (2022)
[13] Wu, F.O., Hassan, A.U., Christodoulides, D.N.: Thermodynamic theory of highly multimoded nonlinear optical systems. Nature Photonics 13(11), 776–782 (2019)
[14] Parto, M., Wu, et al.: Thermodynamic conditions governing the optical temperature and chemical potential in nonlinear highly multimoded photonic systems. Optics letters 44(16), 3936–3939 (2019)
[15] Makris, K.G., Wu, et al.: Statistical mechanics of weakly nonlinear optical multimode gases. Optics letters 45(7), 1651–1654 (2020)
[16] Pathria, R., Beale, P.D.: 1-the statistical basis of thermodynamics. Statistical Mechanics, 1–23 (2011)
[17] Martynov, G.A.: The problem of phase transitions in statistical mechanics. Physics-Uspekhi 42(6), 517 (1999)
[18] Bekenstein, J.D.: Generalized second law of thermodynamics in black-hole physics. Physical Review D 9(12), 3292 (1974)
[19] Marques Muniz, A., Wu, et al.: Observation of photon-photon thermodynamic processes under negative optical temperature conditions. Science 379(6636), 1019–1023 (2023)
[20] Pourbeyram, H., Sidorenko, P., Wu, F.O., Bender, N., Wright, L., Christodoulides, D.N., Wise, F.: Direct observations of thermalization to a Rayleigh-Jeans distribution in multimode optical fibres. Nature Physics 18(6), 685–690 (2022)
[21] Mangini, F., Ferraro, M., Gemechu, W.A., Sun, Y., Gervaziev, M., Kharenko, D., Babin, S., Couderc, V., Wabnitz, S.: On the maximization of entropy in the process of thermalization of highly multimode nonlinear beams. Optics Letters 49(12), 3340–3343 (2024)
[22] Fusaro, A., Garnier, J., Krupa, K., Millot, G., Picozzi, A.: Dramatic acceleration of wave condensation mediated by disorder in multimode fibers. Physical review letters 122(12), 123902 (2019)

[23] Pyrialakos, G.G., Ren, H., Jung, P.S., Khajavikhan, M., Christodoulides, D.N.: Thermalization dynamics of nonlinear non-hermitian optical lattices. Physical Review Letters 128(21), 213901 (2022)
[24] Baudin, K., Garnier, J., Fusaro, A., Berti, N., Michel, C., Krupa, K., Millot, G., Picozzi, A.: Observation of light thermalization to negative-temperature Rayleigh-Jeans equilibrium states in multimode optical fibers. Physical Review Letters 130(6), 063801 (2023)
[25] Ren, H., Pyrialakos, G.G., et al.: Nature of optical thermodynamic pressure exerted in highly multimoded nonlinear systems. Physical Review Letters 131(19), 193802 (2023)
[26] Shi, C., Kottos, T., Shapiro, B.: Controlling optical beam thermalization via band-gap engineering. Physical Review Research 3(3), 033219 (2021)
[27] Ferraro, M., Mangini, F., Zitelli, M., Wabnitz, S.: On spatial beam self-cleaning from the perspective of optical wave thermalization in multimode graded-index fibers. Advances in Physics: X 8(1), 2228018 (2023)
[28] Picozzi, A.: Towards a nonequilibrium thermodynamic description of incoherent nonlinear optics. Optics Express 15(14), 9063–9083 (2007)
[29] Kirsch, M.S., Pyrialakos, G.G., Altenkirch, R., et al.: Observation of Joule–Thomson photon-gas expansion. Nature Physics, 1–7 (2025)
[30] Podivilov, E., Mangini, F., Sidelnikov, O., Ferraro, M., Gervaziev, M., Kharenko, D., Zitelli, M., Fedoruk, M., Babin, S., Wabnitz, S.: Thermalization of orbital angular momentum beams in multimode optical fibers. Physical Review Letters 128(24), 243901 (2022)
[31] Ramos, A., Fernández-Alcázar, L., Kottos, T., Shapiro, B.: Optical phase transitions in photonic networks: a spin-system formulation. Physical Review X 10(3), 031024 (2020)
[32] Weidemann, S., Kremer, M., Helbig, T., Hofmann, T., Stegmaier, A., Greiter, M., Thomale, R., Szameit, A.: Topological funneling of light. Science 368(6488), 311–314 (2020)
[33] Haus, H.A.: Waves and fields in optoelectronics. (No Title) (1984)
[34] Lu, L., Joannopoulos, J.D., Soljačić, M.: Topological photonics. Nature photonics 8(11), 821–829 (2014)
[35] Vakil, A., Engheta, N.: Transformation optics using graphene. Science 332(6035), 1291–1294 (2011)
[36] Plotnik, Y., Peleg, O., Dreisow, F., Heinrich, M., Nolte, S., Szameit, A., Segev, M.: Experimental observation of optical bound states in the continuum. Physical review letters 107(18), 183901 (2011)
[37] Kildishev, A.V., Boltasseva, A., Shalaev, V.M.: Planar photonics with metasurfaces. Science 339(6125), 1232009 (2013)
[38] Hsu, C.W., Zhen, B., Lee, J., Chua, S.-L., Johnson, S.G., Joannopoulos, J.D., Soljačić, M.: Observation of trapped light within the radiation continuum. Nature 499(7457), 188–191 (2013)
[39] Yu, Z., Fan, S.: Complete optical isolation created by indirect interband photonic transitions. Nature photonics 3(2), 91–94 (2009)
[40] Alù, A., Engheta, N.: Achieving transparency with plasmonic and metamaterial coatings. Physical Review E—Statistical, Nonlinear, and Soft Matter Physics 72(1), 016623 (2005)
[41] Bar-Hillel, L., Dikopoltsev, A., Kam, A., Sharabi, Y., Segal, O., Lustig, E., Segev, M.: Time refraction and time reflection above critical angle for total internal reflection. Physical Review Letters 132(26), 263802 (2024)
[42] Brongersma, M.L., Shalaev, V.M.: The case for plasmonics. Science 328(5977), 440–441 (2010)
[43] Lederer, F., Stegeman, G.I., Christodoulides, D.N., Assanto, G., Segev, M., Silberberg, Y.: Discrete solitons in optics. Physics Reports 463(1-3), 1–126 (2008)
[44] Kivshar, Y.S., Campbell, D.K.: Peierls-nabarro potential barrier for highly localized nonlinear modes. Physical Review E 48(4), 3077 (1993)
[45] Stegeman, G.I., Segev, M.: Optical spatial solitons and their interactions: Universality and diversity. Science 286(5444), 1518–1523 (1999)
[46] Peschel, U., Pertsch, T., Lederer, F.: Optical Bloch oscillations in waveguide arrays. Optics letters 23(21), 1701–1703 (1998)
[47] Morandotti, R., Peschel, U., Aitchison, J., Eisenberg, H., Silberberg, Y.: Experimental observation of linear and nonlinear optical Bloch oscillations. Physical Review Letters 83(23), 4756 (1999)

[48] Agrawal, G.P.: Nonlinear fiber optics. In: Nonlinear Science at the Dawn of the 21st Century, pp. 195–211. Springer, (2000)

[49] Chen, H.-H., Liu, C.-S.: Solitons in nonuniform media. Physical Review Letters 37(11), 693 (1976)

**Data availability**

Source data are available for this paper. All other data supporting the plots and findings within this paper are available from the corresponding authors upon request.

**Code availability**

The numerical codes used in this study (MATLAB) are available upon request from the corresponding authors.

**Acknowledgments**

Research partially supported by the U.S. Department of Energy, Office of Basic Energy Sciences, Division of Materials Sciences and Engineering under Award DE-SC0025224 (Developing the theory and building experimental setup) (H.M.D., A.B.B., M.A.S., H.R., G.G.P., D.N.C., and M.K.), Army Research Office (ARO) (W911NF-23-1-0312) (H.M.D., A.B.B., M.A.S., H.R., G.G.P., D.N.C., and M.K.), the Air Force Office of Scientific Research (AFOSR) Multidisciplinary University Research Initiative (MURI) award on Novel light-matter interactions in topologically non-trivial Weyl semimetal structures and systems (award no. FA9550-20-1-0322) (H.M.D., A.B.B., M.A.S., H.R., G.G.P., D.N.C., and M.K.), ONR MURI award on the classical entanglement of light (award no. N00014-20-1-2789) (H.M.D., A.B.B., M.A.S., H.R., G.G.P., D.N.C., and M.K.), AFOSR MURI on Programmable systems with non-Hermitian quantum dynamics (award no. FA9550-21-1-0202) (H.M.D., A.B.B., M.A.S., H.R., G.G.P., D.N.C., and M.K.), Department of Energy grant DESC0022282 (M.A.S., H.R., G.G.P., and D.N.C.), W.M. Keck Foundation (M.A.S., H.R., G.G.P., and D.N.C.), MPS Simons collaboration (Simons grant no. 733682) (M.A.S., H.R., G.G.P., and D.N.C.), and US Air Force Research Laboratory (FA86511820019) (M.A.S., H.R., G.G.P., and D.N.C.).

**Author Contributions**

G.G.P., H.M.D., D.N.C., and M.K. developed the idea. H.M.D. and G.G.P. performed the simulations. H.M.D. and A.M.B.B. built the setup, and H.M.D. performed the experiments. All authors contributed to the analysis of the results and preparation of the manuscript.

**Competing Interests**

The authors declare no competing interests.